\documentclass{acm_proc_article-sp}
\usepackage{booktabs}
\usepackage{graphicx}
\usepackage{placeins}
\usepackage{url}
\begin{document}

\title{Twitter User Classification using Ambient Metadata }

%
%
%
%
%

\numberofauthors{2} 
%
\author{
%
%
\alignauthor Chirag Nagpal\\
       \affaddr{Dept. of Computer Engineering}\\
       \affaddr{Army Institute of Technology}\\
       \affaddr{Pune, India}\\
       \email{chiragnagpal\_12102@aitpune.edu.in}
\alignauthor Khushboo Singhal\\
       \affaddr{Supercomputer Education \& Research Center}\\
       \affaddr{Indian Institute of Science}\\
       \affaddr{Bangalore, India}\\
       \email{khushboo\_singhal@daiict.ac.in}
}
\date{31 July 2014}

\maketitle
\begin{abstract}

Microblogging websites, especially Twitter have become an important means of communication, in todays time. Often these services have been found to be faster than conventional news services. With millions of users, a need was felt to classify users based on ambient metadata associated with their user accounts. We particularly look at the effectiveness of the `profile description' field in order to carry out the task of user classification. Our results show that such metadata can be an effective feature for any classification task.

\end{abstract}




\section{Introduction}

Twitter, over the years has gained immense popularity with milllions of registered users. As a microblogging website, Twitter allows users to post terse 140 character long `tweets' or status updates. For any information retrieval or reccommendation task, user classification can be an effective pre-processing step. Previous work \cite{conf/icwsm/PennacchiottiP11} have utilised profile features and some tweets of a user to bring about the task of user classification, in our study we do \textbf{not} consider the tweets of a user, but utilise other features like followers count, following count, number of tweets and the profile description of a user. The profile description is a short 160 character alphanumeric field. To the best of our knowledge no previous attempt at user classification has utilised this feature. We train two classifiers, OUC, to classify profiles as Organisations, Users and Others and MPS, to classify users on the basis of interest towards Music, Politics and Sports. In order to train the classifier we utilise Decision Trees, Naive Bayes and Support Vector Machines. Comparative results are provided in Section 3. 

\section{Feature Set}

Any classifier, requires a set of features to be trained upon, for our classifier, as mentioned earlier, we utilise a) The followers count, b) The following count, c) The number of Tweets, d) Ratio of Followers to Following count and the e) Profile Description. Features a,b,c are whole numbers whereas the Profile Description is a 160 character alphanumeric field.

\subsubsection*{Profile Description}
We analysed a corpus of over 70,000 user profiles, which were aggregated using the Twitter Streaming API \cite{tdev} and found that about 85\% provide atleast one or more character of user description. We removed all punctuation marks and special characters from all the descriptions. Out of the users that do provide profile description , the average length was found to be 59 characters while the average number of words was approximately 6. 
\begin{figure}[h]
\centering
\includegraphics[width = 6cm]{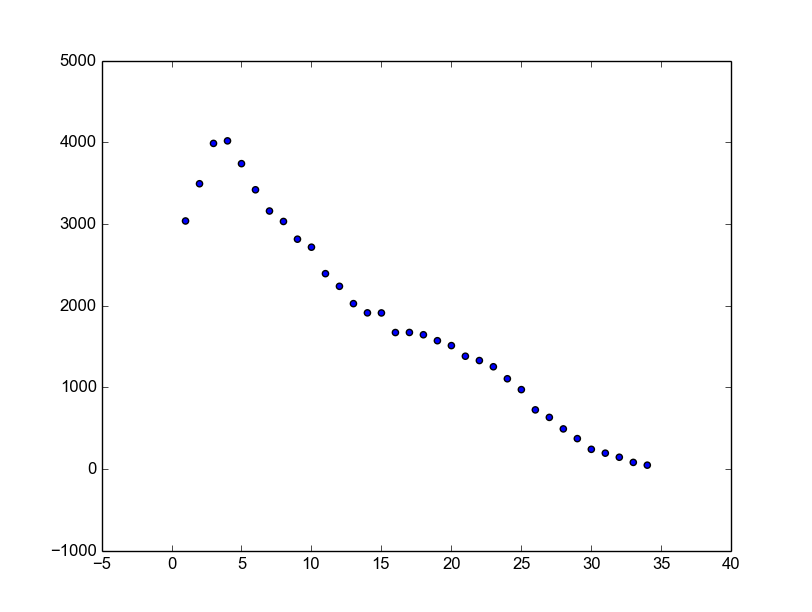}
\caption{No. of users vs. no. of words}
\label{fig:exectelnet}
\end{figure}

We analysed our corpus of user profiles and extracted the top 50 most frequently occuring words in the profile description. These words were used for creating binary features to train our algorithms.
\subsubsection*{Numerical Features}
Since, the range of values for the numerical features was large, binning was required to reduce dimensionality of our features. This was carried out using the function
\begin{equation}\large H(n)=[\log_{10}n]\end{equation}
Here n is the whole number value associated with the particular feature and [ ] is the greatest integer function. Figures 2 to 5 visualise the Features after binning using the function described above.

We also define another derived feature, `ratio' which is essentially the ratio of the followers count and the following count. Since this produces a fractional term, to this we apply the function as described above to perform binning.
\begin{figure}
\centering
\includegraphics[width = 6cm]{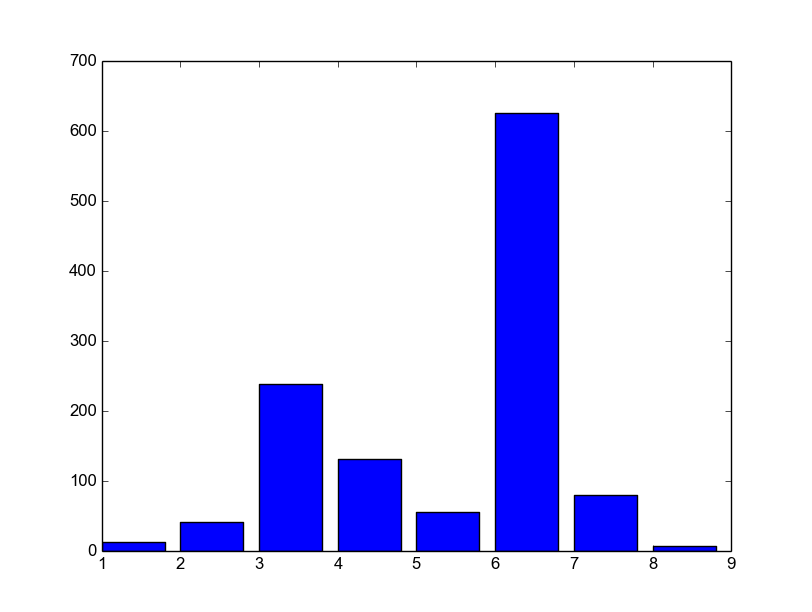}
\caption{No. of Users vs Followers Count}

\includegraphics[width = 6cm]{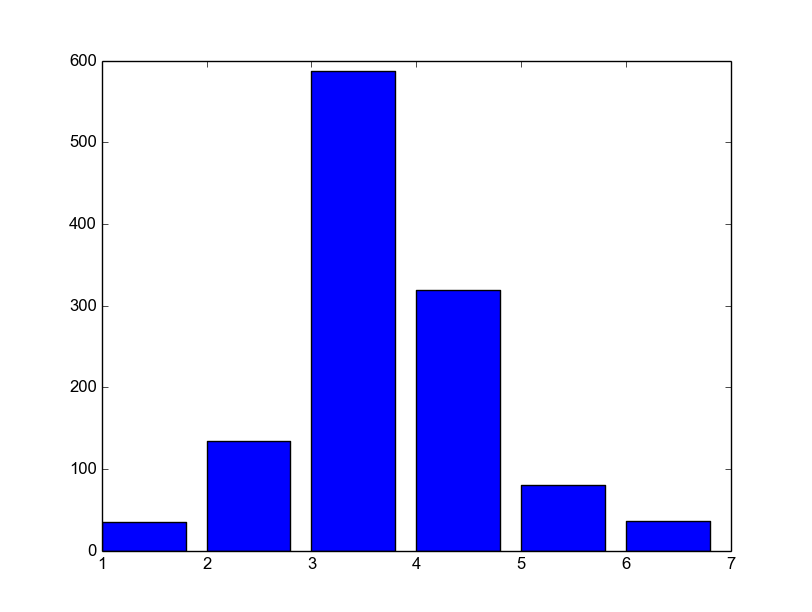}
\caption{No. of Users vs Followings Count}

\includegraphics[width = 6cm]{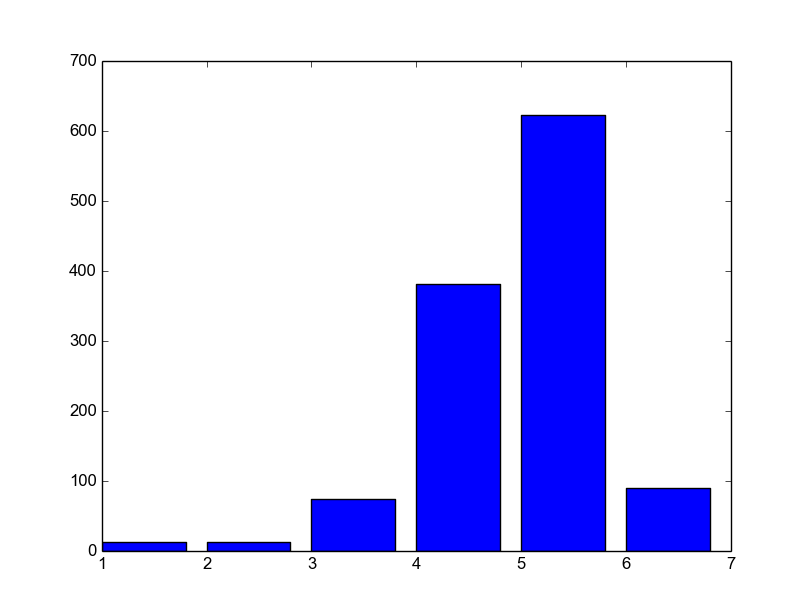}
\caption{No. of Users vs Tweets Count}

\includegraphics[width = 6cm]{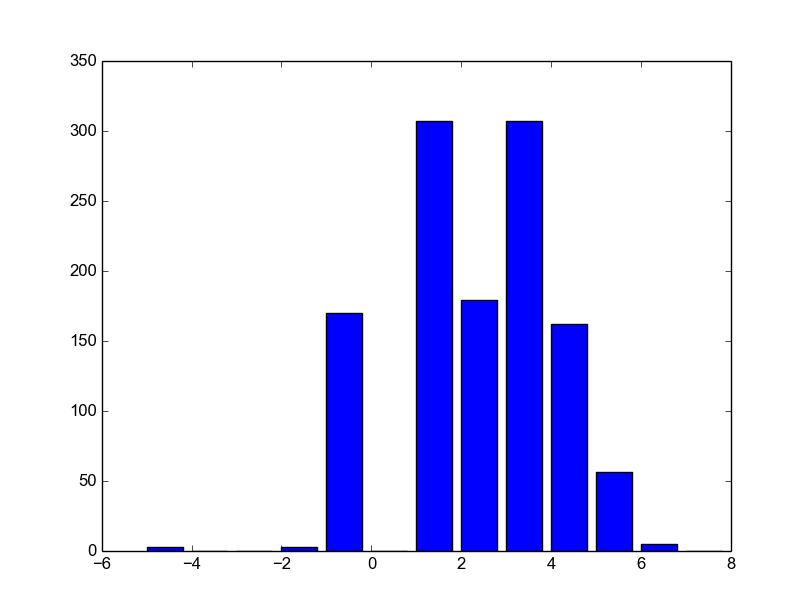}
\caption{No. of Users vs Ratio}
\end{figure}

\section{Training \& Results}

We trained to classifiers, OUC to classify users as Organisations, Users and Celebrites and MPC, Music, Politics and Sports. We aggregated 1200 user profiles and labelled them as OUC and MPS. For each classification we trained Decision Trees (DT), Naive Bayes (NB) and Linear Support Vector Machines (SVM). For DT \& NB we utilise the NLTK Python package \cite{Bird:2009:NLP:1717171}, while for SVM we utilise SciKit \cite{scikit-learn}. 

For each classification we perform 4-cross validation, and present the most accurate confusion matrix out of the four iterations. We also provide the average accuracy of the entire classifier. 
\subsection{OUC Classifier}

\subsubsection{Decision Trees}
\vspace{0.5em}
\begin{table}[h]
\centering

\begin{tabular}{c|ccc}
\toprule \textsc{} & \textsc{U} &
\textsc{O} & \textsc{C} \\
    
\midrule \textsc{U} & \textbf{36.8}\% &  1.9\%         	  &   0.9\% \\
\textsc{O} 			&  	7.5\%		  & \textbf{16.0\%}   &     15.1\%  \\
\textsc{C} 			& 	0.9\%			  &	 1.9\%           &\textbf{18.9\%}	
\\
\bottomrule
\multicolumn{4}{c}{Accuracy: 71.7\%} \\
\end{tabular}
\caption[OUC Classification - Decision Trees]{Confusion Matrix for DT in OUC}
\end{table}
\subsubsection{Support Vector Machines}
\vspace{0.5em}
\begin{table}[h]
\centering
\begin{tabular}{c|ccc}
\toprule \textsc{} & \textsc{U} &
\textsc{O} & \textsc{C} \\
    
\midrule \textsc{U} & \textbf{36.8}\% &  2.8\%         	  &   . \\
\textsc{O} 			&  	7.5\%		  & \textbf{17.9\%}   &     13.2\%  \\
\textsc{C} 			& 	0.9\%			  &	 .           &\textbf{20.8\%}	
\\
\bottomrule
\multicolumn{4}{c}{Accuracy: 75.5\%} \\
\end{tabular}
\caption[OUC Classification - Decision Trees]{Confusion Matrix for SVM in OUC}
\end{table}
\subsubsection{Naive Bayes}
\vspace{0.5em}
\begin{table}[h]
\centering
\begin{tabular}{c|ccc}
\toprule \textsc{} & \textsc{U} &
\textsc{O} & \textsc{C} \\
    
\midrule \textsc{U} & \textbf{37.1}\% 	    &  .         &   1.9\% \\
\textsc{O} 			&  	1.9\%		& \textbf{22.9\%}   &     4.8\%  \\
\textsc{C} 			& 	1.9\%		&	 2.9\%   &\textbf{27.6\%}	   \\
\bottomrule
\multicolumn{4}{c}{Accuracy: 87.6\%} \\
\end{tabular}
\caption[OUC Classification - Decision Trees]{Confusion Matrix for NB in OUC}
\end{table}

\begin{table}[h]

\centering
\begin{tabular}{c|c|c|cc}
\toprule
\textbf{} & \textbf{Feature} & \textbf{Label} & \textbf{Ratio}\\
\midrule 
1&	followers = 6	& c : u 	&34.1 : 1.0	\\
2&	ratio = 4		& c : u 	&26.9 : 1.0 \\
3&  ratio = 3  		& c : u 	&15.6 : 1.0 \\
4& contains(my)		& u : o		&14.1 : 1.0 \\
5& followers = 4	& u : c		&13.6 : 1.0 \\
6& contains(news)	& o : u 	&13.1 : 1.0 \\
7&contains(official)& o : u 	&13.0 : 1.0 \\
8&contains(from)	& o : c		&11.2 : 1.0 \\
9& contains(i) 		& u : o		&9.9 : 1.0  \\
10&followers = 3 	& u : o 	&8.7 : 1.0  \\
\bottomrule
\end{tabular}
\caption[OUC Classification - Most Significant Features]{Most Significant Features in Naive Bayes}
\end{table}

\begin{table}[h]

\centering
\begin{tabular}{c|c|c|cc}

\toprule
\textbf{Features} & \textbf{DT} & \textbf{SVM} & \textbf{NB}\\
\midrule 
numerical		&	65.6\%		&	66.7\%	& 64.8\%	\\
numerical+ratio&		64.6\%	&	65.6\%	&66.3\%	\\
numrical+ratio+description& 69.6\%  & 72.9\%	&80.9\%    \\
\bottomrule

\end{tabular}
\caption[OUC Classification - Overall]{Average  Results for OUC classification}
\end{table}

\newpage

\subsection{MPS Classifier}

\subsubsection{Decision Trees}
\vspace{0.5em}
\begin{table}[h]
\centering
\begin{tabular}{c|cccc}
\toprule \textsc{} & \textsc{M} &
\textsc{P} & \textsc{S}   \\
    
\midrule \textsc{M} & \textbf{30.9\%} & 1.4\% & 2.2\%  \\
\textsc{P} &2.9\%&\textbf{32.5\%}&4.1\%&  \\
\textsc{S} &3.8\% &2.2\% & \textbf{20.1\%}  & \\

\bottomrule
\multicolumn{5}{c}{Accuracy: 83.4\%} \\
\end{tabular}
\caption{Confusion Matrix for DT in MPS}
 \begin{flushleft}

\subsubsection{Naive Bayes}
 \end{flushleft}

\centering
\begin{tabular}{c|cccc}
\toprule \textsc{} & \textsc{M} &
\textsc{P} & \textsc{S}   \\
    
\midrule \textsc{M} & \textbf{23.8\%} & 2.9\% & 7.6\%  \\
\textsc{P} &4.8\%&\textbf{29.5\%}&3.8\%&  \\
\textsc{S} &10.5\% &5.7\% & \textbf{11.4\%}  & \\

\bottomrule
\multicolumn{5}{c}{Accuracy: 76.8\%} \\
\end{tabular}

\caption[OUC Classification - Naive Bayes]{Confusion Matrix for NB in MPS}
\label{tab:misuse-vs-anomaly}
\end{table}

\begin{table}[h]

\centering
\begin{tabular}{c|c|c|cc}
\toprule
\textbf{} & \textbf{Feature} & \textbf{Label} & \textbf{Ratio}\\
\midrule 
1&	contains(music)	& m : p  	&23.4 : 1.0	\\
2&	contains(sports)&  s : p   	&10.6 : 1.0 \\
3&  contains(news)	&  p : m  	&9.4 : 1.0 \\
4& contains(breaking)& p : s		& 8.6 : 1.0
\\
5& contains(new)     &  m : p		& 8.3 : 1.0
 \\
6& contains(it)	&   m : p	&6.7 : 1.0 \\
7&contains(father)&s : p 	& 6.2 : 1.0 \\
8&contains(singer)& m : p		&5.9 : 1.0 \\
9& contains(india) 		&m : p	&5.9 : 1.0  \\
10& contains(by) 	&  p : m 	& 5.3 : 1.0 \\
\bottomrule
\end{tabular}
\caption[MPS Classification - Most Significant Features]{Most Significant Features in Naive Bayes}
\end{table}

\begin{table}[h]

\centering
\begin{tabular}{c|c|c|cc}

\toprule
\textbf{Features} & \textbf{DT} & \textbf{SVM} & \textbf{NB}\\
\midrule 
numerical		&	50.5\%		&	*	& 42.1\%	\\
numerical+ratio&		53.4\%	&	*	&43.1\%	\\
numrical+ratio+description& 80.1\%  & *	&65.3\%    \\
\bottomrule
 
\end{tabular}
\caption[MPS Classication]{Average  Results for MSP classification}
\end{table}

\section{Conclusions}
We find that ambient profile metadata associated with user profiles, is an efficient feature set to bring about user classification. For both classifiers, MPS and OUC we get reasonably high accuracies after 4 cross validation. As expected, from Table 4 we can infer that the OUC classifier depends more on numerical features like followers count whereas, from Table 5 it is clear that the MPS classifier relies more on the lexical features of the profile description. We can thus conclude that any future twitter classifier, must incorporate these features, in its training.

\section{Acknowledgments}
The authors would like to thank Prof. N Balakrishnan for constant guidance and support. We would also like to thank the Supercomputer Education Research Centre, Indian Institute of Science, Bangalore and the Indian Academy of Sciences, The National Academy of Sciences, India and the Indian National Science Academy for providing the opportunity to carry out the research work.
\nocite{*}
%
\bibliographystyle{abbrv}
\bibliography{sigproc}  
%
%


\end{document}